\documentclass[sigconf,authorversion,nonacm]{acmart}
\usepackage{csquotes}

\DeclareMathOperator*{\argmin}{argmin} %

\AtBeginDocument{%
  \providecommand\BibTeX{{%
    \normalfont B\kern-0.5em{\scshape i\kern-0.25em b}\kern-0.8em\TeX}}}

\setcopyright{acmcopyright}
\copyrightyear{2018}
\acmYear{2018}
\acmDOI{XXXXXXX.XXXXXXX}

\newcommand{\sparagraph}[1]{\vspace{1mm} \noindent \textbf{#1} }

\begin{document}

\title{Learned Query Superoptimization}

\author{Ryan Marcus}
\email{rcmarcus@seas.upenn.edu}
\orcid{0000-0002-1279-1124}
\affiliation{%
  \institution{University of Pennsylvania}
  \country{}
}

\begin{abstract}
Traditional query optimizers are designed to be fast and stateless: each query is quickly optimized using approximate statistics, sent off to the execution engine, and promptly forgotten. Recent work on learned query optimization have shown that it is possible for a query optimizer to ``learn from its mistakes,'' correcting erroneous query plans the next time a plan is produced. But what if query optimizers could avoid mistakes entirely? This paper presents the idea of learned query superoptimization. A new generation of query superoptimizers could autonomously experiment to discover optimal plans using exploration-driven algorithms, iterative Bayesian optimization, and program synthesis. While such superoptimizers will take significantly longer to optimize a given query, superoptimizers have the potential to massively accelerate a large number of important repetitive queries being executed on data systems today.
\end{abstract}

\maketitle
\section{Introduction}

Traditional cost-based query optimizers~\cite{systemr} are designed to be fast. When a new query arrives, the optimizer performs some computations using statistical cardinality estimates, and produces a query plan. In this sense, traditional query optimizers  can be viewed as cheap (compared to query execution) stateless functions. Each new query is a blank slate, representing a ``fire and forget'' system.

Past work on learned query optimization~\cite{neo, balsa, rejoin, lero, leo} have addressed the ``forget'' component of ``fire and forget'' optimizers. Since every optimizer processes queries with similar qualities (all querying the same database), one can say there is \emph{cross entropy} (shared information) between each query. By learning a model of query performance that takes advantage of this cross entropy, learned query optimizers ``learn from their mistakes.''  Since learned query optimizers are starting to see early commercial adoption~\cite{bao_scope, bao_scope2}, we can conclude that the database community has made some progress on the ``forget'' element of traditional optimizers.

But what about the ``fire'' element? Query optimizers are built so that optimization time is low compared to query execution time, so the cost of query optimization can be seen as amortized over the query's execution. The simplifying assumption at work here is that since each incoming query \emph{could} be unlike anything seen before, the time spent optimizing any specific query might not be useful for optimizing other queries. When we view each query as arbitrary, this assumption seems valid.

But in many real applications, query workloads are highly repetitive. The same query template -- and often the exact same query --  might be executed many times. In fact, for many analytic dashboarding systems, the \emph{majority of cluster resources} might go to executing a set of highly-repetitive queries (e.g., the hourly sales dashboard executes nearly-identical queries every hour).

Given the shape of these analytics workloads, does it make sense to view each query optimization as a cheap stateless function that only relies on optimizer statistics? This issue has been partially addressed by past work on parameterized or parametric query optimization (PQO)~\cite{kapil_pgo, ppqo, pqo_orig} and query plan caching~\cite{ms_query_store, parametric_plan_cache}, which has mostly focused on further reducing optimization times (as opposed to query latency) by avoiding repetitive work, although there are some exceptions~\cite{kepler}.

Here, I propose \textbf{learned query superoptimization}. What if instead of bounding the query optimizer to a few hundred milliseconds, we instead used hours or even days of computation to optimize a query? For queries that execute thousands or even millions of times a year (as may be the case in large dashboard applications), the additional optimization time could possibly ``pay for itself.''

The thought of using significant resources on query optimization may be viscerally unappealing to many in the database community. To assuage these feelings, note that ``superoptimization'' is a common concept in program compilation: since some programs are executed so often, spending a significant amount of time optimizing a program may be worthwhile even if the program's execution time only improves marginally. If we are willing to spend extra time compiling our DBMSes to eke-out every last drop of performance, why not give DBMS users the same option for their queries?

This paper presents two different directions for future research on learned query superoptimization.

\sparagraph{Exhaustive plan search (Section~\ref{sec:exhaust})} It is well known that cardinality estimates are often poor, and often the cause of poor query plans~\cite{howgood}. While inaccurate, cardinality estimates are fast to compute -- but what if we took a different approach, and designed a query optimizer that intrusively looked at data and executed queries to get around inaccurate cost estimates? 

\sparagraph{Program synthesis (Section~\ref{sec:synthesis})} While DBMSes seek to provide a fast implementation of the relational model to users, the abstractions of the relational model often seep into DBMS design. Bespoke systems (e.g., Millwheel~\cite{millwheel}, Bigtable~\cite{bigtable}) fully discard the generality of the relational model, which allows them to use custom implementations that often outperform anything a general DBMS could offer. Expanding on~\cite{synthesis_layout}, what if program synthesis techniques we could \emph{automatically} generate a bespoke data system for a user? 

To begin, Section~\ref{sec:repetition} will show summary statistics from a large, highly-repetitive workload that motivates the idea of query superoptimization. Then, Section~\ref{sec:exhaust}~and~\ref{sec:synthesis} will discuss future directions for learned query superoptimization work.

\section{Repetition in the real world}
\label{sec:repetition}

\begin{table}
    \centering
    \begin{tabular}{rrrr}
    \toprule
        Duration      & \# templates & \% cluster time & P50 \# executions \\
        \midrule
        $<$ 1 week    & 52         & 3\%   & $< 1000$ \\
        1 - 4 weeks   & 181        & 5\%   & $< 1000$ \\
        4 - 12 weeks  & 1092       & 6\%   & 40900\\
        12 - 24 weeks & 540        & 19\%  & 8700\\
        24 - 52 weeks & 10983      & 31\%  & 108600\\
        \midrule
        Total         & 12848      & 64\%  & $\approx 100000$ \\
        \bottomrule
    \end{tabular}
    \vspace{2mm}
    \caption{The number of query templates executed on a commercial workload, grouped by their lifespan. The P50 (median) number of executions per template is rounded to the nearest thousand. Query templates used for at least half a year to a full year (the last row) represents 31\% of the cluster's total compute time.}
    \label{tab:templates}
\end{table}

In modern analytics systems, repetitive parameterized queries are common. Dashboards, along with weekly and monthly reports, are regularly implemented using parameterized queries.

Of course, modern analytics systems also get a good number of ad-hoc queries. While collaborating with a large corporation, I analyzed a year of query logs from an on-premise data warehouse to determine how often parameterized queries were executed. The results are summarized in Table~\ref{tab:templates}.

There were nearly $11,000$ query templates that appeared in the logs for six months to a full year, and execution of those templates occupied 31\% of the cluster's resources throughout the 12-month period. It is worth noting that these queries are executed quite often: the median query template that lived for six months to one year was executed over a $100,000$ times. With such a large percentage of cluster resources going to executing these repetitive queries, it seems reasonable to try and optimize repetitive queries specifically.

\section{Exhaustive plan search}
\label{sec:exhaust}
For queries joining $n$ relations, optimizers must search $O(3^n)$~\cite{joe_complexity} plans when considering just join ordering, access paths (e.g., index vs. full scan), and operator selection (e.g., hash vs. merge). The key design principles of most traditional optimizers are (1) quickly eliminate large unpromising parts of the plan space, \emph{narrowing} the search space to a manageable size, and (2) use pre-computed \emph{statistics} to find an optimal query plan without significant data processing (i.e., without scanning the underlying data). Unfortunately, these key design principles are also often the root cause of suboptimal query plans:

\sparagraph{A narrow search space} Traditional optimizers must use heuristics to exclude parts of the exponentially-large search space. Unfortunately, these heuristics can exclude the optimal plan. For example, the PostgreSQL optimizer excludes any plan containing cross joins from consideration, but if two dimension tables are sufficiently small, a cross join could be optimal. The reason traditional optimizers exclude cross joins is because plans with cross joins are \emph{almost} always suboptimal, and even when a cross join is optimal, there is normally a near-optimal plan without a cross join. Nevertheless, while unusual, such heuristics can exclude optimal query plans without finding a near-optimal one~\cite{neo}.

\sparagraph{Optimizer statistics} Traditional query optimizers are at the mercy of their cost models and cardinality estimators. While cardinality estimators achieve reasonably good accuracy for table scans, estimator errors often reach catastrophic levels after only a few joins~\cite{howgood}. Thus, the optimizer must plan the final joins of a query plan (which are often the most significant) with virtually no statistics. Even learned query optimization~\cite{rejoin,neo,balsa,lero} follow this paradigm: learned query optimizers still use heuristics to prune the plan space (e.g., the anytime search in~\cite{neo}), and still only operate with pre-computed statistics: the only difference is that the pruning heuristic and statistics are learned.

What if we built a query superoptimizer that ignored both of these fundamental design principles? The simplest possible superoptimizer escaping this paradigm might ``wrap'' a traditional optimizer: execute the top $k$ queries produced by the traditional optimizer and pick the best one, budgeting for the highest $k$ possible. Such an optimizer would verify the quality of $k$ query plans on actual data, albeit in a crude fashion. Another possibility is using an evolutionary algorithm to search for better plans, as in~\cite{kepler}, or to use on-the-fly data sampling for join planning as in~\cite{umbra}.

But we can do much better: it may be possible to build query superoptimizers that interleave machine learning and query execution. Next, I propose two possible systems using this paradigm.

\subsection{Design 1: repeated reinforcement learning}

Reinforcement learning powered optimizers like Neo~\cite{neo}, Balsa~\cite{balsa}, and Lero~\cite{lero} work in \emph{episodes}. Each episode corresponds to building a complete execution plan for a given query. Each episode also ends with a \emph{reward}, the signal that the learned optimizer uses to adjust future actions.

Reinforcement learning algorithms must navigate the \emph{exploration-exploitation tradeoff}. In short, a learning agent must choose to either (1) explore, trying something new and potentially gaining valuable information, or (2) exploit, try something that is known to work already and reap a reward. Existing learned query optimizers seek to perfectly balance exploration and exploitation, maximizing long term rewards~\cite{rl_book}.

A learned query superoptimizer could tilt the scale significantly towards exploration. Doing so would cause the underlying reinforcement learning algorithm to produce more diverse and risky plans. The superoptimizer could then execute each plan, possibly on a sample of the database, and return the plan that works the best. This procedure not only produces a higher quality plan, but this procedure also gathers experience faster, allowing the underling reinforcement learning agent to learn faster.

An additional benefit to a reinforcement-learning powered superoptimizer is that many of the features and safety mechanisms built into current learned systems could be removed. For example, Bao~\cite{bao} provides a number of tools, each using significant resources, to ensure regressions do not occur. With a superoptimizer, no such advanced regression-avoiding tooling is needed, since the empirical quality of the query plan can be observed before optimization finishes.

Tipping the balance towards exploration is far from an optimal solution. Since even a superoptimizer must operate in a limited time budget, choosing exactly which plans to test is an optimization problem on its own. Simply selecting the first $k$ exploratory plans might be far from ideal, since all $k$ plans might explore the same part of the search space. Active learning techniques, like those used in Datafarm~\cite{datafarm}, might be applicable here. This particular problem --- choosing the ideal $k$ explorations before settling on a final, reward-granting decision --- is, to the best of my knowledge, yet unstudied by the database or reinforcement learning community.

\subsection{Design 2: latent space optimization}
\begin{figure*}
    \centering
    \includegraphics[width=0.85\textwidth]{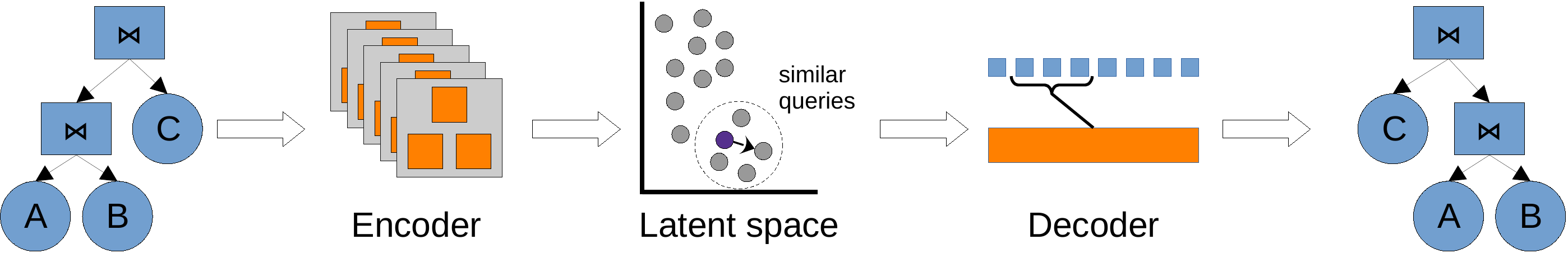}
    \caption{An encoder-decoder design for query optimization. A query plan is ran through a neural network (e.g., tree convolution~\cite{tree_conv}), which transforms the query into a point in a latent space. The encoder is trained to map queries with similar performance properties together. A Bayesian optimizer takes a step in the latent space, moving from the purple circle to the pointed-to circle. The decoder uses generates the new query plan tree, which is executed. The performance of the resulting query plan is given to the Bayesian optimizer, which can then take another step in the latent space.}
    \label{fig:enc_dec}
\end{figure*}

A second possibility is inspired by recent work~\cite{bayes_latent} applying Bayesian techniques to molecular optimization, the task of searching for molecules with particular properties. Translating from finding molecules to finding query plans, the fundamental idea is to \emph{encode} query plans into a \emph{latent space}, use Bayesian optimization techniques to \emph{optimize} the plan within the latent space, and then \emph{decode} the point in the latent space back into a query plan. Figure~\ref{fig:enc_dec} shows an illustration of this process. 

The motivation behind this idea is to translate the problem of query optimization from a ``structural optimization'' problem (i.e., a problem where the output \emph{must} have a particular structure, like a query plan) into a ``continuous optimization'' problem (i.e., a problem in which the output is continuous). Once we have a continuous problem, a large number of well-studied continuous optimization algorithms can be applied (e.g.,~\cite{bayes_local}).

\sparagraph{Encoder \& latent space} The encoder's $E: Q \to \mathbb{R}^n$ goal is to map a query plan $q \in Q$ into a $n$-dimensional latent space $\mathbb{R}^n$ using a neural network. We want positions in the latent space to be semantically relevant, meaning that we want similar points in the latent space to represent similar query plans \emph{and} have similar performance properties (i.e., latency, IOs). One way to do this is with an \emph{information bottleneck}~\cite{bottleneck}: we train a  model to predict the performance properties of a query, but we make one of the last layers (possibly the second or third from the end) intentionally small. This forces the model to learn a compact representation of the query plans in the intentionally small layer, but also requires that the intentionally small layer is organized in such a way that allows for predicting the performance properties of the query. Such a model will not achieve high accuracy compared to models without an information bottleneck, but accuracy is not our goal. We can now chop off the layers after the information bottleneck, and use the resulting network as an encoder: the information bottleneck layer serves as our latent space.

\sparagraph{Decoder} The decoder's $D: \mathbb{R}^n \to Q$ job is to take a point in the latent space created by the encoder and transform the point back into a query plan. The decoder can be trained by differentiating through the encoder, freezing the weights of the encoder during the process. Architecturally, the decoder can be anything that produces a sequence, but a reasonable choice may be a transformer model~\cite{attention} or an LSTM~\cite{lstm}.

\sparagraph{Bayesian optimization} The core of a latent space query superoptimizer is Bayesian optimization. The query optimization problem can be cast as a black-box optimization problem. Let the query plan $p_1$ be the plan generated by a traditional query optimizer, which will be the initial condition for our optimization. Further, let $L(p)$ be the latency of $p$, determined by executing the plan. The Bayesian optimization algorithm then searches for a vector $\hat{v}$ such that:
\begin{equation*}
\underset{\hat{v}}{\argmin} \quad L(D(E(p_1) + \hat{v}))
\end{equation*}

\noindent
The point $E(p_1) + \hat{v}$ can then be decoded into a query plan. 

Bayesian optimization of expensive functions with continuous inputs is a well-studied problem, with many algorithms available~\cite{bayes_survey}. Reinforcement learning powered query optimizers had to invent customized algorithms to deal with the specific challenges of query optimization. By tapping into a preexisting field like Bayesian optimization (which have already had proven success in drug discovery~\cite{bayes_local}, hardware design~\cite{bayes_opt1}, and even urban planning~\cite{bayes_opt2}),  a learned query superoptimizer may be in reach.

\section{Program synthesis}
\label{sec:synthesis}
The previous proposals for query superoptimizers have all assumed that the final output of a query optimizer should be an executable query plan. In this section, we propose thinking more broadly about the potential for a query optimizer that simultaneously optimizes query execution plans \emph{and} the underlying physical layout of the data. To illustrate this potential, imagine writing a program that answers queries about the relationships between actors, movies, and production companies. The system only needs to be able to answer questions in the form of $Q_1$: 
\begin{displayquote}
$Q_1$: How many movies were produced by company $C$ and starred actor $A$?
\end{displayquote}

\begin{figure}
    \centering
    \includegraphics[width=0.35\textwidth]{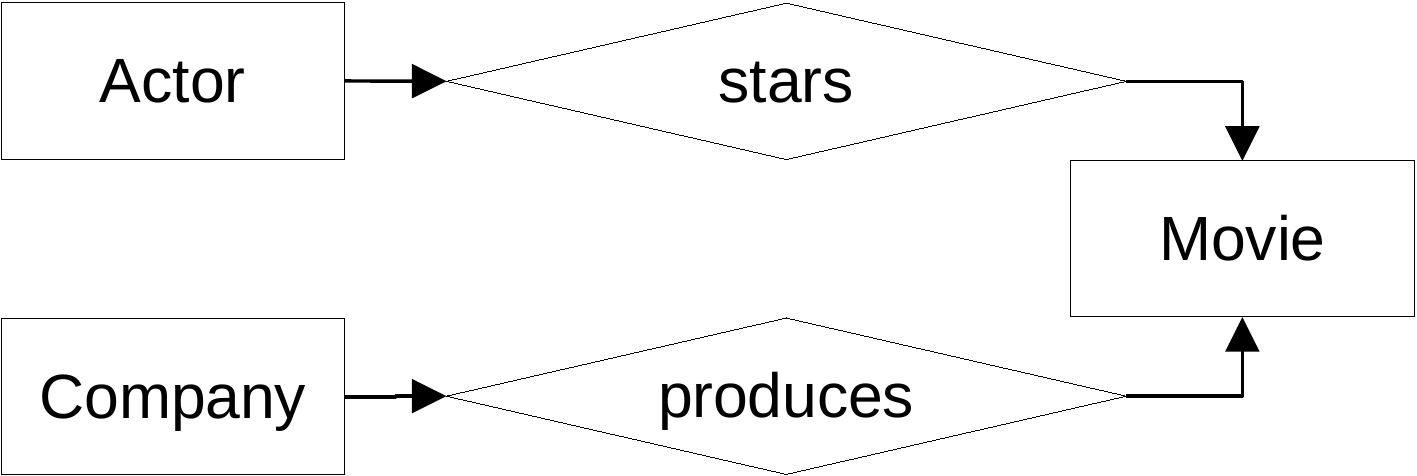}
    \caption{ER diagram of actors, movies, and production companies (attributes omitted). The standard realization into a physical schema would contain five relations.}
    \label{fig:er}
\end{figure}

You could implement this system using a relational database, containing five relations, as shown in the ER diagram in Figure~\ref{fig:er}. Answering the query for any $A$ and $C$ can now be done with a simple SQL query. The underlying optimizer will join the five relations together, possibly even changing the join order based on $A$ and $C$. Note that, because both \texttt{stars} and \texttt{produces} are many-to-many, using fewer than 5 relations would violate second normal form. 

Do we think this implementation is optimal? Of course not. Sure, using a relational database has great engineering benefits, such as (1) the reuse of existing components, (2) the ability to extend the system to other queries, and (3) most organizations already know how to maintain a DBMS. But, if our data system's job is to answer $Q_1$ as fast as possible, then we know that the algorithm executed by the relational database will not be optimal. 

Instead of four joins, imagine we store two hashmaps: one mapping each actor to a bitmap, where that bitmap stores a $1$ if the actor appears in that movie, and a second mapping each company to a bitmap, where that bitmap stores a $1$ if the company produces that movie. Now, answering $Q_1$ is as simple as two hashmap lookups and a bitmap count-intersect. With a cuckoo hashmap and smart prefetching, this could be implemented with only 6 cache misses.

\begin{figure}
    \centering
    \includegraphics[width=0.42\textwidth]{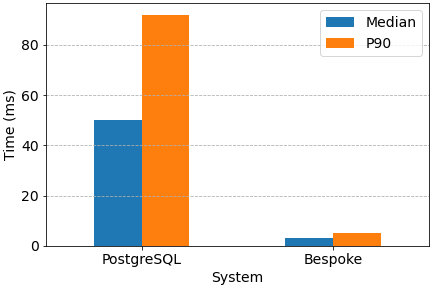}
    \vspace{-3mm}
    \caption{Performance of PostgreSQL vs. the bespoke two-hashmap system for $Q_1$. The bespoke system performs significantly better at both the 50th (median) and 90th percentiles.}
    \label{fig:pg_vs_bespoke}
    \vspace{-4mm}
\end{figure}

Figure~\ref{fig:pg_vs_bespoke} compares the performance of PostgreSQL and an implementation of the bespoke system implemented in Rust using the standard library's hashmap and Roaring~\cite{roaring} bitmaps. The bespoke solution is an order of magnitude faster than the RDBMS. Further, a traditional RDBMS has no chance of replicating the execution strategy of the custom solution.\footnote{One could manually create bitmap indexes and query the count-intersection of them, but this is essentially just creating the customized solution inside of the DBMS instead of allowing the DBMS optimizer to find an optimal plan.}

The above question may seem trivial: a smart DBA could likely create a similar effect using clever materialized views. To see how program synthesis could help in less trivial cases, imagine designing a system to answer $Q_2$:

\begin{displayquote}
$Q_2$: How many movies were produced by $C$ and starred actor $A$ with a rating $R$ s.t. $R_1 < R \leq R_2$.
\end{displayquote}

The rating of a movie can be stored as a computed attribute of \texttt{Movie} in Figure~\ref{fig:er}. Assuming a 5-star review system, a synthesized system similar to the system for $Q_1$ could store a bitmap for each discrete rating (i.e., a bitmap for 1 star films, 2 star films, 3 star films, etc.), and then queries could be answered by counting the number of $R_2$-star movies and subtracting the number of $R_1$-star movies.

What the customized hashmap solution gains in speed, it loses in generality. Adding additional query types would  require a significant re-write. Implementing transaction semantics for adding new actors, movies, or companies may be challenging. But, we know that bespoke data management systems are occasionally created for important applications (e.g., Google Ads, YouTube), often at a high cost. Depending on the business value of answering $Q_1$ or $Q_2$ quickly, a more bespoke solution may be worthwhile.

How can we make DBMSes capable of \emph{automatically} generating the customized hashmap solution for answering $Q_1$ or $Q_2$, while simultaneously providing the generality and ease-of-use of SQL? Even capturing 50\% of the bespoke system's performance in a generic RDBMS would be a huge win.

One possible way to make that happen is \emph{machine programming}~\cite{pillars}, or getting computers to program themselves. Imagine that we can map out the myriad of potential optimizations, from bitmap intersections to augmented binary trees. It is possible that program synthesis techniques, taking only the schema and SQL query as an input, could be used to construct provably-correct, custom-tailored data systems like the hashmap solution described above. Combined with the right learning technique, a synthesis approach might be able to create data structures and algorithms that even outperform human experts.

Building a mapping or library of potential optimizations may seem challenging, but there are many reasons to hope. Data Calculator~\cite{data_calculator} showed how different data structures can be composed together in a provably-correct way to build a variety of key-value stores. Castor~\cite{synthesis_layout} makes progress in terms of defining a formal language that could potentially express more complex data layouts. GenesisDB~\cite{genesisdb} showed how large language models might be able to create customized database components.

\section{Related work}
Query optimization is one of the most well-studied problems in the database community, with a history spanning 50 years~\cite{systemr}. More recently, we have seen applications of machine learning techniques~\cite{deep_card_est2, zeroshot_latency_model, learn_cost, deep_card_est_koudas, zero_shot_streaming}, especially reinforcement learning~\cite{neo, bao, balsa, lero, qo_rank}, to the problem of query optimization. \cite{synthesis_layout, data_calculator}~show how program synthesis techniques can be used to create custom-tailored relational layouts or key-value stores.

In general, machine programming~\cite{pillars} represents a distinct field of research, and includes work about garbage collection~\cite{learned_gc}, static analysis~\cite{controlflag}, and program regression detection~\cite{perf_reg_zeroshot}. Bayesian optimization is a well-studied area, and~\cite{bayes_survey}~provides a survey of this area. \cite{bayes_latent}~inspired the latent space optimization proposed here. The idea of autonomous experimentation on top of a database system has also been previously explored~\cite{ergalics, ergalics_db}. To the best of my knowledge, the term ``superoptimization'' was coined in~\cite{superopt_coined}.

\section{Conclusions}
This paper has proposed learned query superoptimization, an opportunity for the database community to redefine the task of query optimization for increasingly-common repetitive analytic workloads. Many of the ideas mentioned here will be pursued in the coming years. Collaborators welcome!

\clearpage

\bibliographystyle{ACM-Reference-Format}
\bibliography{ryan-cites-long}

\clearpage
\end{document}